\documentstyle[12pt]{article}
\input epsf
\catcode`@=11
\def\jimtwoup{\twocolumn\sloppy\flushbottom\parindent 2em
        \parskip .33\baselineskip
        \leftmargini 2em\leftmarginv .5em\leftmarginvi .5em
        \oddsidemargin 0in      \evensidemargin 0in
        \columnsep .4in \footheight 0pt
        \textwidth 10in \topmargin  -.4in
        \headheight 0pt \topskip 0in
        \textheight 6.9in \footskip 30pt
        \hoffset -.5in \voffset -.25in
        \def\@oddfoot{\hfil\thepage\hfil\addtocounter{page}{1}
                \hspace{\columnsep}\hfil\thepage\hfil}
        \let\@evenfoot\@oddfoot \def\@oddhead{} \def\@evenhead{} }
\catcode`@=12

\begin{document}

\title{ Two Ising Models Coupled to 2--Dimensional Gravity}
\author{
Mark Bowick, Marco Falcioni, \\
Geoffrey Harris and Enzo Marinari$^{(*)}$\\[1.5em]
Dept. of Physics and NPAC,\\
Syracuse University,\\
Syracuse, NY 13244-1130, USA\\
{\footnotesize
  bowick
  falcioni
  gharris
  @npac.syr.edu}\\
{\footnotesize
  marinari@roma1.infn.it}\\[1.0em]
{\small $(*)$:  and Dipartimento di Fisica and INFN,} \\
{\small Universit\`a di Roma {\it Tor Vergata}}\\
{\small Viale della Ricerca Scientifica, 00173 Roma, Italy}}
\maketitle

\begin{abstract}
To investigate the properties of $c=1$ matter coupled to
$2$d{--}gravity we have performed large-scale simulations of two
copies of the Ising Model on a dynamical lattice.  We measure spin
susceptibility and percolation critical exponents using finite-size
scaling.  We show explicitly how logarithmic corrections are needed
for a proper comparison with theoretical exponents.  We also exhibit
correlations, mediated by gravity, between the energy and magnetic
properties of the two Ising species.  The prospects for extending this
work beyond $c=1$ are addressed.
\end{abstract}

\begin{flushright}
  {SU-HEP-93-4241-556}\\
  {SCCS 540}\\
  {\bf hep-th/9310136}\\
\end{flushright}

\newpage

\section{Introduction \protect\label{S_INT} }

There is at present considerable analytic understanding of how
conformal matter with central charge ($c$) less than or equal to one
couples to two-dimensional gravity \cite{DFGIZJ}.  These $c \leq 1$
models are relevant both as non-critical string theories and as novel
statistical mechanical systems describing matter on a dynamical
substrate lattice.  The relation of these models to quantum Liouville
theory \cite{KPZ} allows one to compute the shift in the scaling
dimension of physical operators due to the coupling to 2d-gravity.
This shift is determined solely by the scaling dimension of the
operator in the absence of gravity (i.e on a fixed lattice) and the
central charge of the theory.  Given the dressed scaling dimensions
one can compute critical exponents of thermodynamic observables
related to correlation functions.  Through the mapping of these
systems onto matrix models it is even possible to incorporate topology
change by performing the entire sum over topologies in the so-called
double-scaling limit.
This beautiful state of affairs falls apart for central charge
exceeding one\footnote {More precisely when the quantity $c - 24
\Delta > 1$, where $\Delta$ is the conformal weight of the
lowest-weight state in the theory
\cite{KUSE}.}.
In this case the methods of Liouville theory appear to fail - in
particular they predict complex critical exponents. The vanishing of
the mass of the dressed identity operator (the tachyon) at $c=1$
suggests the onset of an instability in the worldsheet geometry.
Based on this, it is widely believed that $c >1$ no longer describes a
continuum theory of surfaces \cite{GinMoo}.  Since no models have been
solved in this regime, it would be valuable to apply numerical
techniques to see if $c >1$ models appear to be qualitatively
different than those with $c < 1$.

A simple class of models to test our understanding of these issues is
provided by multiple copies of the Ising model on a dynamical lattice.
Each Ising model has $c = \frac{1}{2}$. A model with $n$ individual
copies then has $c = n/2$. For a single Ising model ($n=1$) on a
dynamical lattice the critical exponents may be computed analytically
via a mapping of the model onto a particular 2-matrix model (with one
matrix representing spin up and the other spin down)
\cite {KAZ,BOKA}. The model has a third order rather than a second order
phase transition.  Without the coupling to gravity the partition
function of the $n$-Ising model would simply be the $n${--}th power of
the single Ising model partition function, since the spins of
different copies are independent.  The interaction with gravity,
however, induces an effective interaction between spins of different
species (copies). Spins of different species effectively interact
through the dynamical lattice.  Br\'ezin and Hikami \cite
{BRHI,HIBR,HIK} have studied an equivalent $2^n$-matrix model
realization of this system in perturbation theory in the cosmological
constant, without any apparent anomaly setting in at $c=1$. Similarly
Monte-Carlo studies of multiple{--}Ising and multiple{--}Potts models
on dynamically-triangulated random surfaces (DTRS) do not uncover
dramatic changes as $c$ passes through one \cite{BJ,CKR,AMBETAL}.

On the numerical side it is important that we understand the
transition case $c = 1$ before plunging into the regime $c>1$.  This
is the motivation for the work presented in this paper on large-scale
DTRS simulations of the $n=1$ and $n=2$ Ising model coupled to
2d-gravity.  In the $n=2$ case the formalism of KPZ allows a
computation of the relevant critical exponents, which we can
consequently compare to the numerical results extracted from
finite-size scaling and direct fits.  The exact solution of the
single-Ising model also provides a direct check that the discrete DTRS
algorithm is reproducing continuum behavior for the lattice sizes
simulated and that the numerical analysis methods employed are
adequate. The $n=2$ ($c=1$) case is considerably more complicated than
the $n=1$ case because of logarithmic violations of scaling. Here the
extensive work on $c=1$ matrix models provides an essential clue
\cite {KLE}.
It allows us to identify the appropriate scaling variable that
replaces the cosmological constant. With this in hand we are able to
compare the results of our large-scale DTRS Monte-Carlo simulations
with the predictions of KPZ.

      Similar issues are also addressed in a companion paper
\cite{percpaper} in which percolation coupled to gravity with
$c=0,1/2,1$ and $c > 1$ matter is examined in much detail.  The nature
of finite size effects in simulations of two dimensional gravity plays
a central role in that paper, as it does here.  We shall show how the
percolation results for Gaussian matter in \cite{percpaper} are
consistent with those obtained in the two species Ising model
discussed here.

The outline of the rest of the paper is as follows.  In section 2 we
give a theoretical discussion of $c=1$ conformal matter coupled to
2d-gravity, with a derivation of the critical exponents we will be
comparing with our numerical simulations and a discussion of
logarithmic violations of scaling.  In section 3 we present our
numerical methods and results including a comparison with theoretical
expectations.  Finally, in section 4, we conclude with a discussion of
the origin of logarithmic corrections to scaling and the prospects for
extending this work to the regime $c > 1$ that this implies.


\section{Theoretical Predictions}
\label{theory}
       We shall consider a model in which Ising spins are attached to
the vertices of triangulations.  The triangulations are characterized
by their adjacency matrix $C_{ij}$ which equals $1$ if $i$ and $j$ are
neighbors and vanishes otherwise.  $C_{ij}$ is the discrete analogue
of the worldsheet metric $g_{ij}$.  We shall restrict ourselves to the
set of triangulations with $N$ vertices ${\cal{T}}_N$ containing only
loops of length $3$ or greater and vertices of coordination number of
at least $3$.  We simulate a theory determined by the partition
function
\begin{equation}
     Z_N = \sum_{T\in{\cal{T}}_N} \sum_{\sigma_i =  \pm 1}
\exp ( -\beta
\sum_{\alpha=1}^{n_s} \sum_{i,j=1}
^{N} C_{ij}(T)\sigma_i^{\alpha}\sigma_j^{\alpha});
\protect\label{partfn}
\end{equation}
$\alpha$ labels the spin species.  In this paper, we address the
cases $n_s = 1$ and $2$.
 Most of the relevant theoretical calculations are performed
in the grand-canonical ensemble, with the partition function
\begin{equation}
     Z(\mu) = \sum_{N=1}^{\infty} Z_N \exp (-\mu N)
\end{equation}
dependent on $\mu$, the cosmological constant.

      Our primary observable will be the spin--susceptibility, which we
express as the integrated spin--spin correlation function
\begin{equation}
{\chi}_N = \frac{1}{n_s~N}~\langle \sum_{\alpha=1}^{n_s}\sum_{i,j}
\sigma_i^{\alpha}
\sigma_j^{\alpha} \rangle.
\end{equation}
The integrated spin--spin correlation function in the grand--canonical
ensemble then satisfies
\begin{equation}
\protect\label{suscgcan}
      \langle \sum_{\alpha = 1}^{n_s}\sum_{i,j}
\sigma_i^{\alpha} \sigma_j^{\alpha}
\rangle (\mu) Z(\mu) = \sum_{N=1}^{\infty} n_s N\chi_NZ_N \exp (-\mu N).
\end{equation}

Using standard arguments \cite{DDK} one can determine how the scaling
behavior of the integrated spin--spin correlation function changes
under coupling to gravity.  In flat space, the spin--spin correlation
functions scales as
\begin{equation}
\protect\label{spinscale}
 \langle \sigma_i^{\alpha}\sigma_j^{\alpha} \rangle \sim
|{\vec{r}}_i - {\vec{r}}_j|^{-2(\Delta_{\sigma}^{o} + {\bar{\Delta}}_
{\sigma}^{o})}.
\end{equation}
The weight $\Delta_{\sigma}^{o} = {\bar{\Delta}}_{\sigma}^{o}
 = 1/16$
is dressed by gravity in a theory
of central charge $c$ according to the KPZ formula \cite{KPZ}
\begin{equation}
\protect\label{KPZmsz}
(\Delta_{\sigma} - \Delta_{\sigma}^{o}) = \left(1 +
{\frac{1}{12}}\left(\sqrt{1-c} - \sqrt{25-c}~\right)\sqrt{1-c}
\right)\Delta_{\sigma}
(1 - \Delta_{\sigma}).
\end{equation}
      This dressed weight determines the scaling of the integrated
spin--spin correlation function with $\mu$ on surfaces of genus $h$:
\begin{equation}
\protect\label{suscmusca}
      \langle \sum_{\alpha} \sum_{i,j} \sigma_i^{\alpha} \sigma_j^{\alpha}
\rangle  (\mu)  Z(\mu) \sim (\mu - \mu_c)^{2(-1 + \Delta_{\sigma}
) + (2-{\gamma}_s)(1-h)}
\end{equation}
(the $-1$ before the dressed weight
accounts for the integrations of $i$ and $j$ over the surface) with
\begin{equation}
     \gamma_s = {\frac{1}{12}}\left( c - 1 - {\sqrt{(25-c)(1-c)}}\right).
\end{equation}
The above relations  and
\begin{equation}
Z_N \sim N^{-1 + (\gamma_s - 2)(1-h)}
\end{equation}
yield the finite-size scaling relation
\begin{equation}
\protect\label{meanszscal}
  \chi_N \sim N^{\gamma/\nu d_H},
\end{equation}
with
\begin{equation}
\protect\label{gammreln}
\gamma/\nu d_H = 1 - 2\Delta_{\sigma}.
\end{equation}
This is the scaling law that we shall verify numerically.
The susceptibility scales as $\chi \sim (\beta-\beta_c)^{-\gamma}$,
the correlation length (governed by the decay of the spin--spin
correlation function) obeys $\xi \sim (\beta-\beta_c)^{-\nu}$ and
$d_H$ is the intrinsic Hausdorff dimension of the random surface being
considered.

     Assuming the standard scaling hyperscaling relation $\alpha = 2 -
\nu d_H$, we can also predict the value of $\gamma$. The specific heat
scales as $N^{\alpha / \nu d_H}$.  Then by applying the reasoning used
to arrive at (\ref{gammreln}) to the two-point function of the energy
operator $\varepsilon$, one finds $\alpha/ \nu d_H$ equals $1 - 2
\Delta_{\varepsilon}$, where $\Delta_{\varepsilon}$ is the dressed
weight of the energy operator.  It then follows that
\begin{equation}
\protect\label{gammadef}
\gamma = {\frac{(1 - 2\Delta_{\sigma})}{(1 - \Delta_{\varepsilon})}}.
\end{equation}
  One obtains $\Delta_{\varepsilon}$
through the KPZ formula (\ref{KPZmsz}), substituting the bare
energy weight $\Delta_{\varepsilon}^{o} = 1/2$ for $\Delta_{\sigma}^{o}$.

   We shall also measure scaling properties of the Fortuin-Kasteleyn
(FK) clusters {\cite{FK}} which we construct to update the spin
degrees of freedom.  The FK clusters appear in the reformulation of
the Ising model as a correlated spin--bond percolation model with
partition function \cite{Sokal}
\begin{equation}
\protect\label{corrperc}
Z = \sum_{\sigma_i = \pm 1}\sum_{colorings}p^{b}(1-p)^{N_b-b}.
\end{equation}
Colorings consist of a set of `black' bonds drawn between adjacent
points with identical spin values; each black bond is drawn with
probability $p= 1 - \exp (-2\beta)$.  In (\ref{corrperc}) $b$ bonds
out of a possible $N_b$ bonds of the lattice are colored black.  FK
clusters comprise sets of sites linked together by black bonds.
Therefore each cluster is assigned a single spin value.  In the
multi-generation case, we build a set of colored bonds and FK clusters
separately for each species of spin.  One can show that the spin--spin
correlation function of the spins $\langle \sigma_i^{\alpha}
\sigma_j^{\alpha} \rangle $ equals the pair-connectedness function
$\langle \delta_{{\cal{C}}^{\alpha}_i, {\cal{C}}^{\alpha}_j} \rangle$
of the corresponding FK clusters; $\delta$ is $1$ when $i$ and $j$ lie
in the same cluster ${\cal{C}}^{\alpha}$ and $0$ otherwise \cite{Hu}.
{}From this, it follows that the spin--susceptibility equals the mean
cluster size ${\cal{S}} = \langle s^2 \rangle / \langle s \rangle$, in
which averages are taken over the distribution $n(s)$, the mean number
of clusters per configuration containing $s$ sites.  We thus shall
determine the scaling of the mean cluster size and in addition, the
fractal dimension $d_f$ of the largest cluster.  The average maximal
size cluster ${\cal{M}}$ of each configuration scales as
\begin{equation}
{\cal{M}} \sim  N^{\frac{d_f}{d_H}}.
\end{equation}
 Standard scaling arguments \cite{Staufferbook}
relate $\gamma/ \nu d_H = 2d_f/d_H - 1$.  One can derive this, for
instance, by considering the asymptotic form of $n(s) \sim
Ns^{-\tau}$.  The standard hyperscaling relation $\nu d_H = 2 -
\alpha$ is then needed.  The singularity of the cluster number
density, the zeroth moment of $n(s)$, as a function of $p-p_c$, is
given by the exponent $(2-\alpha)$.  Similarly the second moment of
$n(s)$ scales as $(p-p_c)^{-\gamma}$. Thus the usual scaling
assumptions imply $\gamma/\nu d_H = (\tau - 3)/(1 - \tau)$.
${\cal{M}}$ asymptotically obeys
\begin{equation}
     N \int_{\cal{M}}^{N} s^{-\tau} \sim 1
\end{equation}
(that is, the mean number of clusters per configuration of size
greater than ${\cal{M}}$ is of order unity) and hence $d_f/d_H =
1/(\tau - 1)$.  Eliminating $\tau$ then gives the above relation
between $d_f/d_H$ and $\gamma/\nu d_H$.

      We can also obtain additional information about the critical
geometry of these theories by examining the properties of pure
percolation clusters.  Consider the bond-percolation model
\begin{equation}
Z = \sum_{colorings}p^{b}(1-p)^{N_b - b}q^{N_c},
\protect\label{qPottspart}
\end{equation}
which for $q=2$ is yet another formulation of the Ising partition
function and more generally is the partition function of the $q$-state
Potts model.  The pair-connectedness function then exhibits the
scaling behavior at criticality of (\ref{spinscale}) with weights
\cite{DeNijs}
\begin{equation}
\Delta_{\sigma,q}^{o} = {\bar{\Delta}}_{\sigma,q}^{o}  =
\frac{(1 - y^2)}{8(2-y)}; ~~
\cos({\frac{\pi y}{2}}) = {\frac{1}{2}}\sqrt{q}.
\end{equation}
The $q \rightarrow 1$ limit corresponds to pure percolation, which has
no dynamics (and vanishing central charge) and thus does not induce
any back-reaction when it is coupled to a theory of gravity and matter
of central charge $c$.  $\Delta_{\sigma,q=1}^{o} = 5/96$; the dressed
weight is again governed by the KPZ formula
(\ref{KPZmsz}).  With this new dressed weight, we can then predict the
scaling behavior of the mean (and maximal) cluster sizes
${\cal{S}}_{N,q=1}$ (and ${\cal{M}}_{N,q=1}$) using (\ref{meanszscal})
with ${\cal{S}}$ substituted for $\chi$.

      In our simulations, we shall consider site percolation, in which
sites (rather than bonds) are colored black with probability $p$ and
clusters are built by connecting adjacent colored sites.  It is well
known that bond and site percolation are in the same universality
class, so that the scaling predictions described above should still
hold in the case of pure site percolation.  It is advantageous to
consider site percolation because the critical value of $p$, $p_c$, is
constrained to equal $1/2$ for site-percolation on triangulations
\cite{SykesEss}\footnote{There are some possible exceptions to this
constraint, which definitely do not apply in the cases we shall
consider here.  This issue is discussed extensively in
\cite{percpaper}.}.

\subsection{$c=1$}

   The scaling relations become more complicated for $c=1$.  Analytic
solutions of the $c=1$ matrix models (and a careful analysis of
Liouville theory) show that correlation functions no longer scale
simply as powers of the cosmological constant $\mu$.  Instead, the
appropriate scaling variable is $\eta$ which satisfies {\footnote{
Without loss of generality, we set $\mu_c = 0$.}}
\begin{equation}
\protect\label{logscaling}
\mu = -\eta \ln (\eta) + c_1\eta + \cdots
\end{equation}
in the limit of small $\eta$; $c_1$ is a constant that we do not
specify and shall not assume to be universal.  This scaling relation
has been derived for the Gaussian theory (of finite and infinite
radius) coupled to gravity.  The product of Ising models lies on the
Gaussian $c=1$ orbifold line \cite{GinLesHou}; coupled to gravity, it
is not equivalent to the solved matrix models.  We shall assume that
the asymptotic logarithmic scaling violation is characteristic of
$c=1$ and thus holds in the two-species case.  Then we conjecture that
the scaling relation (\ref{suscmusca}) should be modified so that
\footnote{The essential role of
logarithmic corrections in interpreting numerical measurements of
$\gamma_s$ at $c=1$ has been previously discussed in reference
\cite{Ambgs}.}
\begin{equation}
\protect\label{suscetasca}
      \langle \sum_{i,j} \sigma_i^{\alpha} \sigma_j^{\alpha}
\rangle (\mu) Z(\mu) \sim \eta(\mu)^{2(-1 + \Delta_{\sigma}
) + (2-{\gamma}_s)(1-h)}.
\end{equation}
In the case of pure percolation, the pair-connectedness function
should be substituted for the spin--spin correlation function, and the
dressed Ising weight $\Delta_{\sigma} = 1/4$ should be replaced by
$\Delta_{\sigma,q=1} = \sqrt{5/96}$.  In the following formulae,
$\chi$ and the mean cluster size ${\cal{S}}$ are interchangeable.

    Our simulations will be done on worldsheets of toroidal topology
($h=1$) for which $Z(\mu) \sim \ln (\eta)$ for $c=1$ (for those models
that have been solved analytically, so again we are making an
assumption about universality).  To extract the asymptotic scaling
behavior of $\chi_N$, we invert the relation between $\eta$ and $\mu$
order by order in $1/\ln \mu$ and $\ln (-\ln (\mu))/\ln (\mu)$ to
obtain
\begin{equation}
\protect\label{inversion}
\eta = -{\frac{\mu}{\ln \mu}}\left( 1 + {\frac{\ln (-\ln \mu)}{\ln \mu}}
+ \left({\frac{\ln (-\ln \mu)}{\ln \mu}}\right)^2 - {\frac{\ln (-\ln
\mu)}{(\ln \mu)^2}} + \cdots \right) .
\end{equation}
We then expand the inverse Laplace transform of (\ref{suscetasca}) to
obtain
\begin{eqnarray}
\protect\label{susczn}
N\chi_NZ_N \sim {\frac{1}{N(N\ln N)^{\omega}}}\left( 1 - {\frac{\ln
\ln N}{\ln N}} + \left(\frac{\ln \ln N}{\ln N}\right)^2 - \frac{\ln
\ln N}{ (\ln N)^2} + \cdots \right)^{\omega} \times \nonumber \\
\left( 1 + {\frac{\omega \Psi(-\omega)}{\ln N}} -
\omega\Psi(-\omega){\frac{\ln \ln N}{(\ln N)^2}} + \cdots \right);
\end{eqnarray}
$\omega = 2(-1 + \Delta_{\sigma}) = -\gamma/\nu d_H - 1$ (or $\omega =
2(-1 + \Delta_{\sigma,q=1})$ for pure site percolation) and $\Psi$ is
the digamma function.  The scaling behavior of $NZ_N$ is obtained by
inverse Laplace transforming $\partial Z(\eta(\mu))/\partial \mu$:
\begin{equation}
\protect\label{partc1}
Z_N \sim {\frac{1}{N}}\left(1 + {\frac{1}{\ln N}} - {\frac{\ln \ln
N}{(\ln N)^2}} + \cdots \right).
\end{equation}
In addition to the higher order terms that we have dropped from the
inversion, there are additional corrections to the above formulae.
The correction that depends on $c_1$ in (\ref{logscaling}), which we
neglect, should lead to contributions to (\ref{susczn}) and
(\ref{partc1}) that are competitive with the smallest corrections that
we have included above.  In addition, there should be the usual
corrections to the scaling (\ref{suscetasca}), but these will be
suppressed by powers of $1/N$ and are negligible for moderately large
$N$.

      The difference in the theoretically predicted scaling behavior
of percolation clusters at $c=1$, compared to $c=1/2$, illustrates the
sensitivity of the worldsheet geometry to the presence of the Ising
spins.  This should induce an effective coupling between species,
which we should be able to detect through correlations between
different species' observables.  To look for this coupling, we shall
measure
\begin{equation}
\protect\label{enecorr}
e_{\alpha \beta} = {\frac{1}{3N}}\left(\langle E^{\alpha} E^{\beta}
\rangle - \langle E^{\alpha} \rangle \langle E^{\beta} \rangle\right)
\end{equation}
and
\begin{equation}
\protect\label{magcorr}
m_{\alpha \beta} = {\frac{1}{N}}\left( \langle |M^{\alpha}||M^{\beta}|
\rangle - \langle |M^{\alpha}| \rangle \langle |M^{\beta}| \rangle \right)
\end{equation}
with $E^{\alpha} =
\sum_{ij}C_{ij}(T)\sigma_i^{\alpha}\sigma_j^{\alpha}$ and
$|M^{\alpha}| = \sum_i |\sigma_i^{\alpha}|$.  A nonzero effective
coupling between species should then be manifest through the
quantities $e^{*} = 2e_{12}/$tr $e$ and $m^{*} = 2m_{12}/$tr $m$,
which vanish when $E^{\alpha}$ ($|M^{\alpha}|$) are uncorrelated and
are $1$ or $-1$ when they are respectively perfectly correlated or
anti-correlated.

\section{ Numerical Simulations and Results }
\label{numsec}

The partition function (\ref{partfn}) is evaluated numerically by a
Monte Carlo simulation. The sum over triangulations is implemented via
the standard DTRS algorithm \cite{DTRS}. This updates the connectivity
matrix $C_{ij}$ by flips on randomly chosen pairs of triangles sharing
a common link.  The Ising spins are updated using the Swendsen-Wang
algorithm
\cite{SwenWang};
one builds FK clusters over the entire lattice and then assigns a
randomly chosen value to the spin associated with each cluster.

      Runs were performed on lattices of toroidal topology of size $N
= 2048, 4096, 8192$ and $16384$.  Each sweep consisted of $3N$ flips
of randomly chosen links followed by a Swendsen-Wang update of the
spins.  To perform the percolation measurements the sites on the
lattice were then randomly colored and percolation clusters
constructed.  We used the jacknife technique to estimate our errors.
We measured auto-correlation functions and computed integrated auto-
correlation times, using standard techniques {\cite{Sokal}}.  The
magnetization, susceptibility and cluster sizes exhibited considerable
critical slowing down \footnote{More detail on critical slowing down
for various algorithms applied to spin models on random lattices will
appear in future work \cite{csdpaper}.}.  It was necessary therefore
to sample quite a large number of lattices - for each data point
between $15,000$ and $30,000$ independent lattices, requiring from
$30,000$ to $900,000$ sweeps.  The total CPU-time used was
approximately equivalent to six months on an HP-9000 (720)
workstation.  We histogrammed our data \cite{histogrammers}.  For
cluster data on the larger lattices, however, histogramming was not
reliable, given our statistics.  The cluster data exhibited extremely
large fluctuations from one measurement to the next; presumably this
was the source of the poor performance of histogramming.

\begin{figure}
\epsfxsize=4.5in \epsfbox{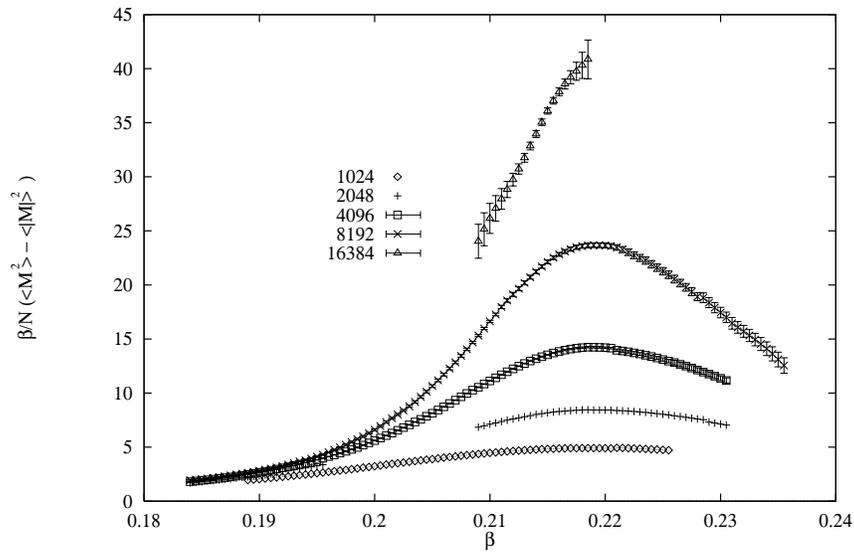}
\protect\caption{\protect\label{suscpeakc12}
The `lattice susceptibility' plotted for the one-species model for
lattices of size $1024-16384$.  Error bars are only shown when they
are larger than the symbols.}
\end{figure}

      To extract critical exponents using finite-size scaling we must
first determine the critical point.  We estimated the value of
$\beta_c$ by locating the peak in the lattice susceptibility $ \beta
/N(\langle M^2
\rangle -
\langle |M| \rangle ^2)$, where $M$ is the magnetization averaged
over species \footnote{Note that this does not equal the
susceptibility $\chi_N$ defined in section (\ref{theory}).  $\chi_N$
contains no subtractions and agrees with the continuum susceptibility
only in the high-temperature phase.}. We first present our results for
a single species.  Figure \ref{suscpeakc12} reveals that this quantity
peaks at a value of $\beta_c = .2185 (20)$.
  We also plotted Binder's cumulants \cite{Binder} ${\cal{U}}_M = 1 -
\langle M^4 \rangle /(3 \langle M^2 \rangle ^2 )$ as a function of
$\beta$ for different lattice sizes.
\begin{figure}
\epsfxsize=4.5in \epsfbox{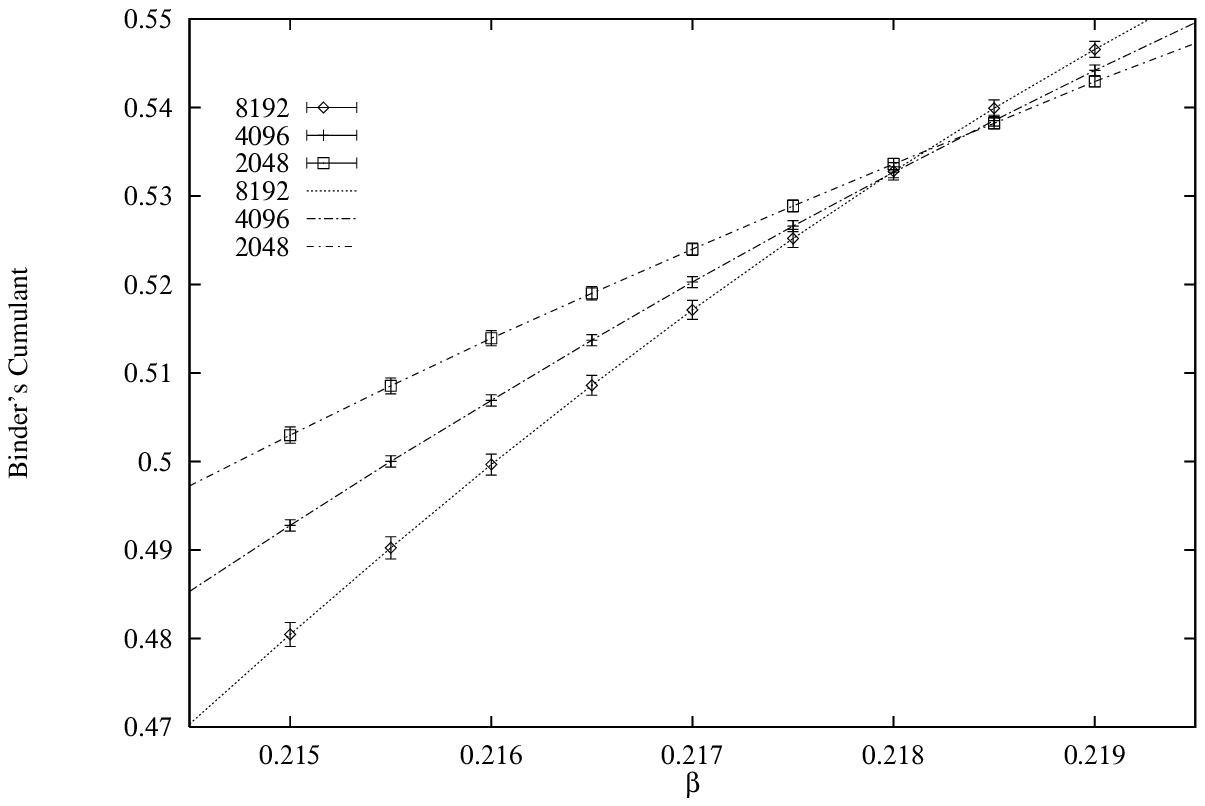}
\protect\caption{\protect\label{bindc12}
Binder's cumulants ${\cal{U}}_M$ for $N=2048,4096$ and $8192$ lattices
in simulations of the one species Ising model for lattices of size
$1024-16384$. Lines are drawn to guide the eye.}
\end{figure}
   From the position of the intersections of these cumulants, we
estimate $\beta_c = .2180 (7)$.  The critical temperature for Ising
spins on lattices dual to our triangulations was computed analytically
in \cite{BurdJur}, so by the Ising duality relation, it follows that
$\beta_c = (1/2)\ln(131/85)
\sim .216273$.  Our numerical estimate of the critical value of $\beta$
is therefore somewhat high.

\begin{figure}
\epsfxsize=4.5in \epsfbox{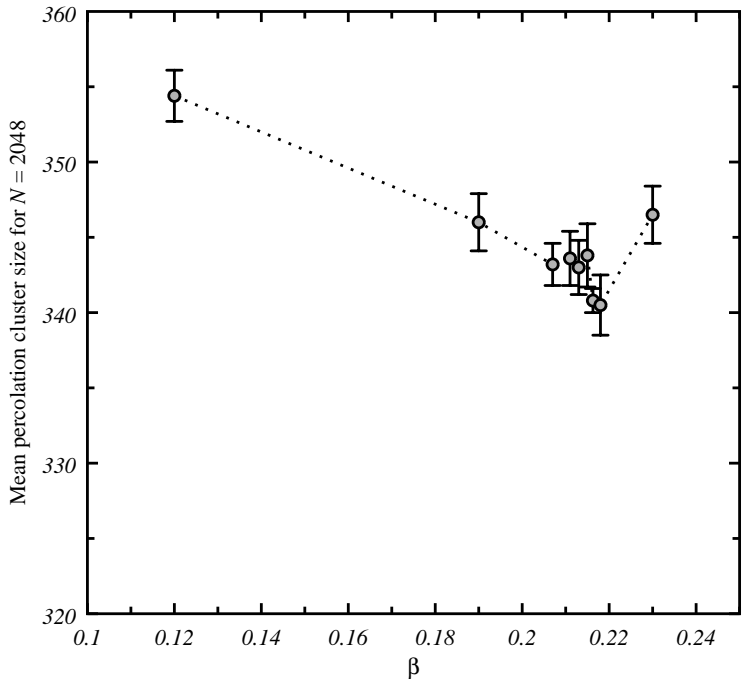}
\protect\caption{\protect\label{mszdip}
The mean size of percolation clusters as a function of $\beta$ for the
one species Ising model on $2048$ node lattices.  The dip occurs near
criticality.  The dotted line is drawn to guide the eye.}
\end{figure}

     One could also envision locating the critical temperature by
looking for a minimum of the mean pure percolation cluster size as a
function of $\beta$.  The growth in the mean cluster size is
determined by the exponent $(\gamma /\nu d_H)_{q=1}$, which decreases
with $c$.  We would therefore expect that percolation clusters for a
given lattice size will become smaller as $c$ increases or likewise as
$\beta$ is tuned to bring the Ising spins to criticality.  In
Fig.~\ref{mszdip} we do indeed observe a dip in the mean cluster size
around the estimated value of $\beta_c$.
The dip is quite broad, however, and doesn't pinpoint $\beta_c$; the
distinction between the $\beta = .218$ and $.210$ points in the figure
may be a statistical artifact.  For $N=4096$, for instance, with
better statistics we find that the mean size is measured to be $571.8
\pm 1.9, 571.1 \pm 1.9, 573.1 \pm 1.9, 572.0 \pm 1.8$ and $572.6 \pm
1.9$ for $ \beta = .211,.213,.215,.216273$ and the $.218$.  Thus we
observe no significant variation over a range $\delta\beta = .007$.
On the other hand the breadth of the dip means that measurements of
$(\gamma/\nu d_H)_{q=1}$ are not very sensitive to the estimated value
of $\beta_c$.  This is not true for the exponent $(\gamma / \nu d_H)$
which, as we shall see, exhibits a strong dependence on the estimated
critical temperature.

       From the finite-size scaling relation (\ref{meanszscal}) we
define the effective exponents
\begin{equation}
    (\gamma/\nu d_H)_{eff} \equiv
\ln\left({\frac{\chi_{2N}}{\chi_{N}}}\right)/
\ln 2.
\end{equation}
When cluster sizes rather than the spin--susceptibility are considered,
it is implicit that ${\cal{S}}$ is to be substituted for $\chi$ in the
above definition.  Likewise,
\begin{equation}
    (d_f/d_H)_{eff} \equiv
\ln\left({\frac{{\cal{M}}_{2N}}{{\cal{M}}_{N}}}\right)
/\ln 2.
\end{equation}
As usual, the analogous exponents for pure percolation are defined as
above with subscript $q=1$.  We now summarize our results for these
critical exponents, extracted through finite size scaling of $\chi_N$,
the mean FK cluster size ${\cal{S}} _{FK}$, the maximal FK cluster
size ${\cal{M}}_{FK}$, the mean percolation cluster size
${\cal{S}}_{q=1}$ and the maximal percolation cluster size
${\cal{M}}_{q=1}$.  In table (\ref{c12exactexp}), we present these
values obtained from runs at the known value of $\beta_c \sim
.216273$.  The exact values are $\gamma/\nu d_H = 2/3$, $d_f/d_H =
5/6$, $(\gamma/ \nu d_H)_{q=1} = (4 - \sqrt{7/2})/3$ and
$(d_f/d_H)_{q=1} = (7 - \sqrt{7/2})/6$.
\begin{table}
\begin{tabular}{| c | c | c | c | c | c |} \hline
$N$ & 1024 & 2048 & 4096 & 8192 & theory \\ \hline $(\gamma/\nu
d_H)_{eff}$ from $\chi$ &.702 (6)& .705 (5)& .686 (6)& .688 (10) & 2/3
\\ \hline $(\gamma/\nu d_H)_{eff}$ from ${\cal{S}}_{FK}$ && .701 (7)&
.690 (9)& .676 (12)& 2/3 \\ \hline $(d_f/d_H)_{eff}$ &&.844 (5)& .839
(7)& .830 (8) & 5/6 \\ \hline $(\gamma/\nu d)^{eff}_{q=1}$ &&.747
(6)&.741 (7)&.722 (10)& .710 \\ \hline $(d_f/d_H)^{eff}_{q=1}$ &&.866
(5) &.866 (6)&.852 (8)& .855 \\ \hline
\end{tabular}
\caption{Summary of exponents extracted from finite-size scaling
at the exact value of $\beta_c$ for one Ising species.  The exponents
are computed using data from lattices of size $N$ and $2N$.}
\label{c12exactexp}
\end{table}
The agreement between our measurements and the theoretical predictions
is quite good.  Since $(d_f/d_H)$ only describes properties of the
largest cluster, it might be less subject to corrections to scaling
(for small cluster sizes) and thus provide the best estimator of the
critical exponents.  Indeed, the values of $(d_f/d_H)_{eff}$ for FK
and percolation clusters on the larger lattices already match the
asymptotic predictions, within our statistics.  The exponents
$(\gamma/\nu d_H)_{eff}$ decrease towards their asymptotic values and
differ from them by only about $2\%$ on the largest lattices.

      In the two species case we do not, however, know $\beta_c$.  To
use the single species model as a control, therefore, we estimate the
critical exponents using finite-size scaling at an estimated value of
$\beta_c = .2180 (7)$.  We present these estimates in table
\ref{c12estexp}.
\begin{table}
\begin{tabular}{| c | c | c | c | c | c |} \hline
$N$ & 1024 & 2048 & 4096 & 8192 & theory \\ \hline $(\gamma/\nu
d_H)_{eff}$ from $\chi$ &.74 (3)& .76 (3)& .73 (2)& .76 (4)& 2/3 \\
\hline $(\gamma/\nu d_H)_{eff}$ from ${\cal{S}}_{FK}$ &&.748 (14) &
.741 (8)&& 2/3 \\ \hline $(d_f/d_H)_{eff}$ &&.877 (15)&.869 (6)&& 5/6
\\ \hline
$(\gamma/\nu d)_{eff,q=1}$ &&.750 (10)& .739 (8) & & .710 \\ \hline
$(d_f/d_H)_{eff,q=1}$ && .871 (7) & .867 (5)& & .855 \\ \hline
\end{tabular}
\caption{Summary of exponents extracted from finite-size scaling
using the numerically determined value of $\beta_c$ for one Ising
species.}
\label{c12estexp}
\end{table}
The values quoted from the scaling of $\chi$ characterize the range of
exponents in the window $\beta \in [.2173, .2187]$; this determination
relies on histogramming.  The exponents extracted from cluster data
are taken from data at $\beta = .218$ (within this window), since
histogramming was not reliable for this data.  We see that the shift
in temperature away from the exact $\beta_c$ induces a large change in
the Ising exponents.  In particular, $(\gamma/\nu d_H)_{eff}$, which
previously came within 2\% of the asymptotic value on the largest
lattice, now is 10--15\% higher.  The percolation exponents fare much
better; they are quite insensitive to the value of $\beta$ and agree
with the values taken at $\beta_c = .216273$.

     We also attempted to perform a direct fit of $\chi$ to $(\beta -
\beta_c)^{-\gamma}$.  We found that a power law fit only seemed to
work in the region $\beta = .19 - .195$, where finite size effects
were not severe.  Only two of our data points (corresponding to the
lowest values of $\beta$ at which we ran) were then used in this fit,
yielding $\gamma = 1.8 (1) $, which is not so far from the theoretical
value of $\gamma = 2$ that follows from (\ref{gammadef}).  Given the
sparsity of our data in this regime and thus the difficulty in
verifying that we are seeing asymptotic scaling, these results should
be interpreted with caution.

     We now turn to a treatment of the two species results.  First, we
present plots of $\beta /N(\langle M^2 \rangle - \langle |M|
\rangle^2)$ and the intersections of Binder's cumulants.

\begin{figure}
\epsfxsize=4.5in \epsfbox{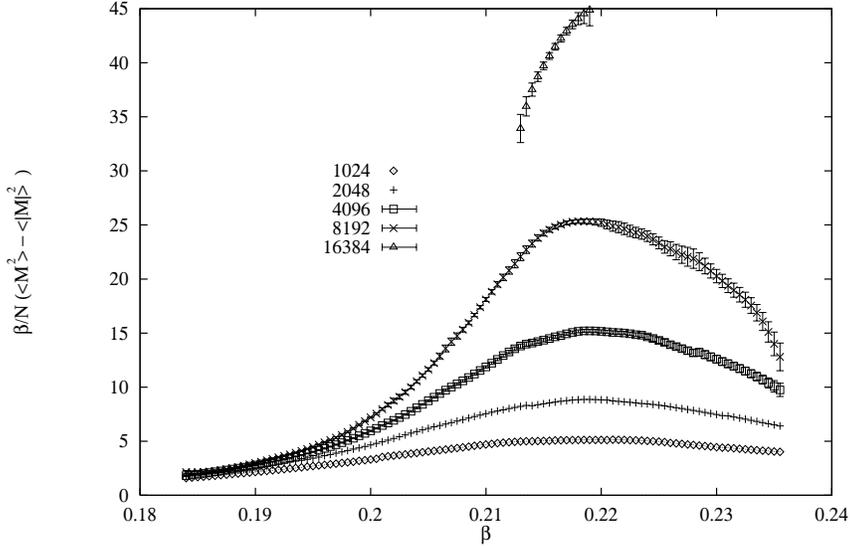}
\protect\caption{\protect\label{suscpeakc1}
A plot of the `lattice susceptibility' for the two species case with
$N=2048,4096,8192$ and $16384$. Error bars are only shown when they
are larger than the symbols.}
\end{figure}
Note that these observables look qualitatively very similar to their
one species counterparts.  The position of the susceptibility peak is
$\beta = .2185^{+35}_{-20}$.  The Binder's cumulants (taken from the
largest lattices where we have sufficient data to determine
intersections) give $\beta_c = .217 (1)$.
\begin{figure}
\epsfxsize=4.5in \epsfbox{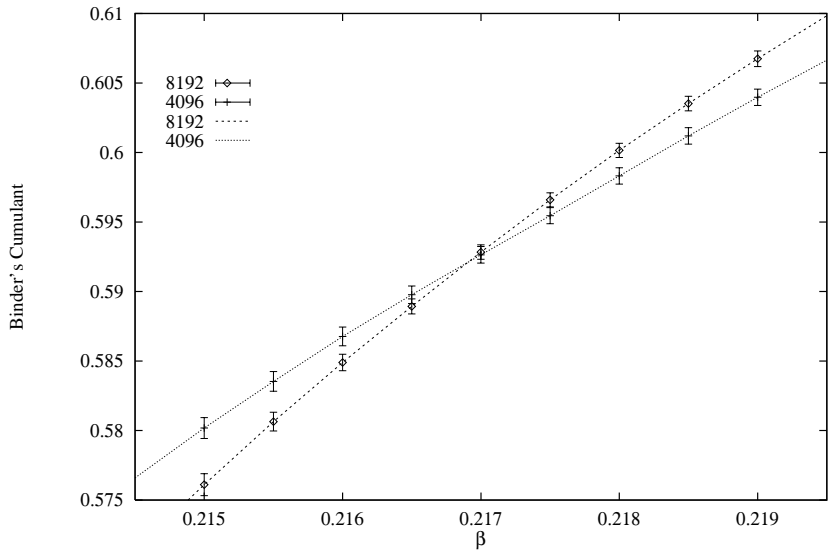}
\protect\caption{\protect\label{bindc1}
Binder's cumulants for $N=4096$ and $N=8192$ as
measured in the two species model.}
\end{figure}
As in the one-species case we adopt the Binder's estimate, which seems
to be more precise.  Based on the logarithmic corrections to scaling
discussed in section \ref{theory}, we would estimate that on the
lattice sizes simulated we are further from the asymptotic scaling
regime than in the one species case.  It is therefore likely that our
estimate of $\beta_c$ here is less precise.

In table~\ref{c1estexp}, we compare the finite-size scaling
measurements of $(\gamma /\nu d_H)_{eff}$ and $(\gamma /\nu d_H)
^{eff}_{q=1}$ with the theoretical exponents, computed from the KPZ
formula, neglecting logarithmic corrections.
\begin{table}
\begin{tabular}{| c | c | c | c | c | c |} \hline
& 1024 & 2048 & 4096 & 8192 & theory \\ \hline
$(\gamma/\nu d)_{eff}$ from $\chi$ &.73 (3)&.73 (3)&.75 (5)&.74 (4)& 1/2
\\ \hline
$(\gamma/\nu d)_{eff,q=1}$ &.759 (11)&.717 (5) &.712 (7) & .709 (10) & .544
 \\ \hline
$(d_f/d_H)_{eff,q=1}$ &.852 (6)&.848 (4) &.848 (5)&.848 (6) & .772 \\ \hline
$(\gamma/\nu d)_{eff,q=1} (X)$ &.739 (7)&.732 (7) &.700 (10) & & .544
 \\ \hline
$(d_f/d_H)_{eff,q=1} (X)$ &.861 (5)&.859 (7) &.841 (8)& & .772 \\ \hline
\end{tabular}
\caption{Summary of exponents extracted from finite-size scaling
for two Ising species compared with KPZ exponents, without logarithmic
corrections.  In the last two rows, we present the
corresponding percolation exponents for lattices on which
a Gaussian field $X$ was simulated.
 }
\label{c1estexp}
\end{table}
The first row contains an estimate of the susceptibility exponent
based on histogram extrapolations within the region $\beta = .216 $ to
$.218$.  We observed that measuring the mean-size of FK clusters (at
$\beta = .216$) always yielded exponents agreeing with those extracted
from the spin--susceptibility.  Since the cluster data did not
histogram reliably, it was difficult though to estimate the
corresponding exponents throughout the above range of $\beta$ without
taking a large amount of additional data.  FK cluster scaling was
primarily measured just to verify that it agreed with the scaling of
the spin--susceptibility.  Our data already showed this, so it did not
seem worthwhile to repeat runs to measure the cluster scaling at
various temperatures.  The following rows thus summarize data for pure
percolation clusters, which was relatively insensitive to $\beta$;
i.e. a shift of $\beta$ of $.005$ did not induce a stastically
significant change in the $q=1$ exponents.  Since again percolation
cluster data did not histogram reliably, we present data taken at
$\beta = .216$ in table~\ref{c1estexp}; these should be representative
of the values one would measure throughout the region $\beta \in
[.216,.218]$.  The final two rows include exponents extracted from
simulations of a Gaussian field $X$ coupled to gravity, as discussed
in \cite{percpaper}.

\begin{figure}
\epsfxsize=4.5in \epsfbox{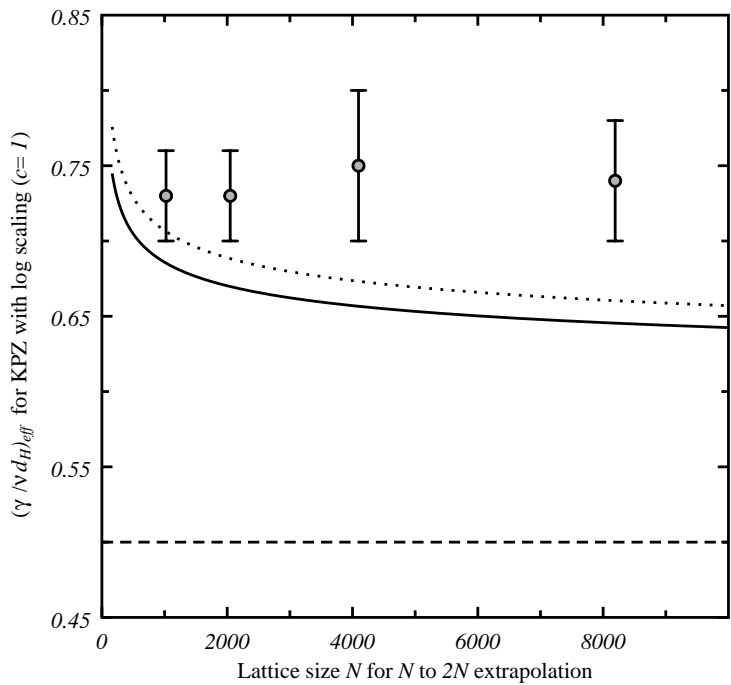}
\protect\caption{\protect\label{suscc1}
A comparison of the theoretical prediction of the finite--size $c=1$
magnetic susceptibility scaling with our data.  The dotted line
includes the leading logarithmic correction and the solid line takes
into account all subleading terms that we calculated. The horizontal
dashed line represents the prediction without logarithmic
corrections.}
\end{figure}

   What is most striking about the measured $c=1$ exponents is in fact
that they agree quite well with the $c=1/2$ data (taken from the
numerically estimated range of $\beta_c$) and in the case of
percolation, the $c = 1/2$ theoretical predictions.  There is clearly,
however, a very large discrepancy between the measured and theoretical
$c=1$ scaling exponents.  On the larger lattices, the percolation
exponents for the two species Ising model also agree fairly well with
those measured in the Gaussian field simulations, suggesting some
degree of universality at $c=1$ in their behavior.  From equations
(\ref{susczn}) and (\ref{partc1}), we can determine the logarithmic
corrections to the exponents $(\gamma /\nu d_H)_{eff}$ and $(\gamma
/\nu d_H)^{eff}_{q=1}$.  One can easily see that these are
considerable by including the leading correction, which gives
\begin{equation}
\protect\label{leadcorr}
   (\gamma /\nu d_H)_{eff} = \gamma /\nu d_H + {\frac{1 + \gamma/
\nu d_H}{\ln N}} + \cdots;
\end{equation}
this formula of course holds for $q=1$.  In Fig.~\ref{suscc1} we
compare the measured $(\gamma /\nu d_H)_{eff}$ with the corresponding
theoretical predictions including both the leading correction
(\ref{leadcorr}) and all of the logarithmic corrections that follow
from equations (\ref{susczn}) and (\ref{partc1}).
The data does not agree particularly well with any of the predictions,
but that is to be expected, since our estimate of $\beta_c$ most
likely induces a large error.  Note that the theoretical curve
including logs lies very close to the $c=1/2$ asymptotic value of
$\gamma /\nu d_H = 2/3$ for the lattice sizes we simulate.  The
measured values of this exponent for $c=1$ and $c=1/2$ (when we use
the numerically estimated values of $\beta_c$) are essentially
identical, so that the discrepancy between the theoretical predictions
(incorporating logs in the $c=1$ case) and the data are thus roughly
the same for $c=1$ and $c=1/2$.

\begin{figure}
\epsfxsize=4.5in \epsfbox{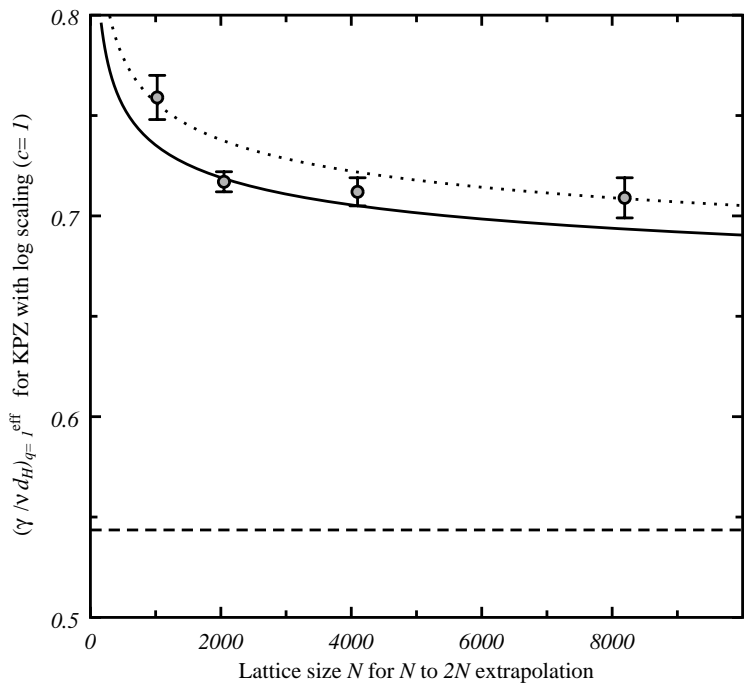}
\protect\caption{\protect\label{suscc1q1}
A comparison of the theoretical prediction of the finite size $c=1$
scaling of $(\gamma/\nu d_H)_{q=1}$ with our percolation data for the
two species Ising model.  The dotted line includes the leading
logarithmic correction and the solid line takes into account all
subleading terms that we calculated. The horizontal dashed line
indicates where these curves asymptote.}
\end{figure}

In Fig.~\ref{suscc1q1} a similar comparison is shown for the exponent
$(\gamma / \nu d_H)_{q=1}^{eff}$.  We see that the data and
theoretical predictions match quite well.
Presumably, as in the $c=1/2$ case, the comparison works because this
exponent no longer depends sensitively on our determination of
$\beta_c$.  The agreement with theory is about as successful as in the
Gaussian case \cite{percpaper}.  This suggests the likelihood that at
least the leading logarithmic corrections to scaling (\ref{leadcorr})
are correct and universal for $c=1$.  We should note here that
although the data can be fit by including the above logarithmic
corrections it would not be possible to extract these corrections from
the data itself without theoretical guidance.

    As in the one species case, we also fitted our lowest $\beta$ data
points to $\chi \sim (\beta - \beta_c)^{-\gamma}$.  The fit yielded
$\gamma = 2.03 (4)$ which does not match the theoretical value of $
\gamma = 1 + 1/\sqrt{2} \sim 1.71$.  It is not clear that this fit is
reliable; probably the exact relation between $\chi$ and $\beta$
should also include logarithmic corrections.

\begin{figure}
\epsfxsize=4.5in \epsfbox{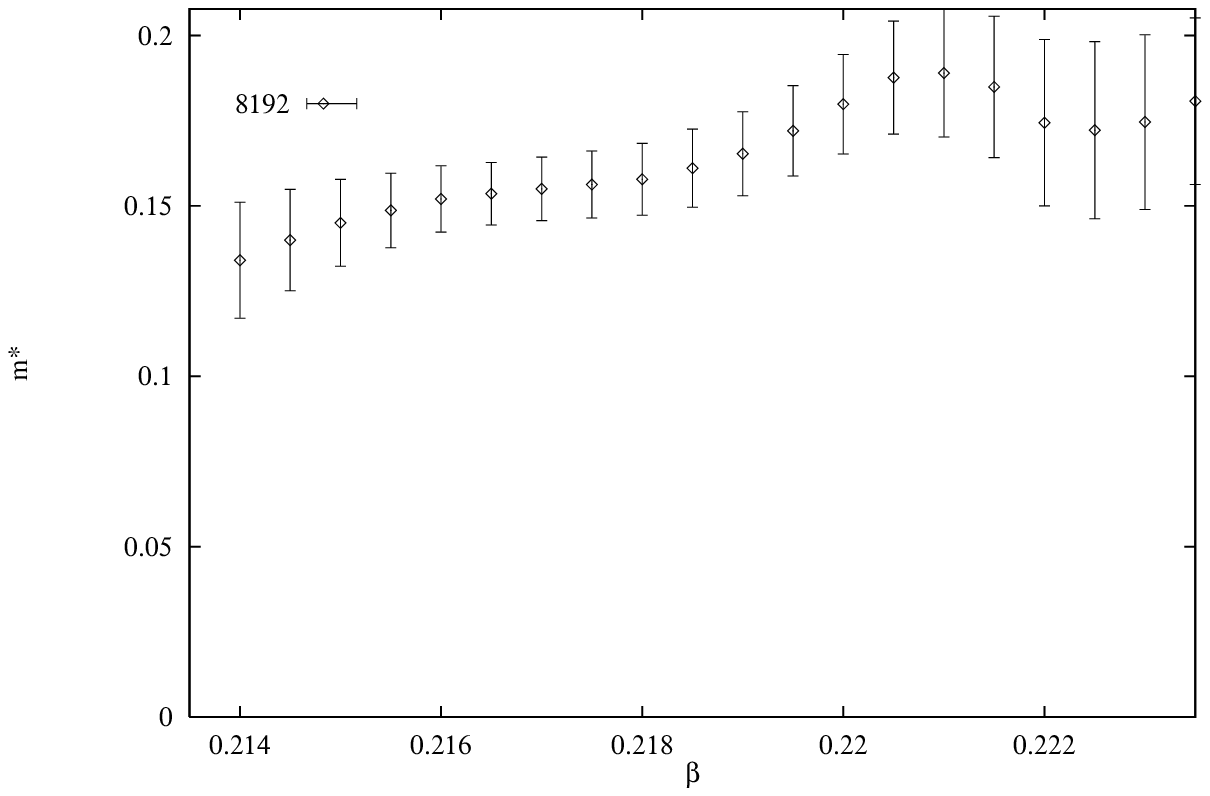}
\protect\caption{\protect\label{magcorrfig}
The correlation $m^*$ plotted for the two species model on $N=8192$
lattices.  Recall that the correlations are normalized so that $m^*$
equals one when $|M^{\alpha=1}|$ and $|M^{\alpha=2}|$ are fully
correlated.}
\end{figure}

\begin{figure}
\epsfxsize=4.5in \epsfbox{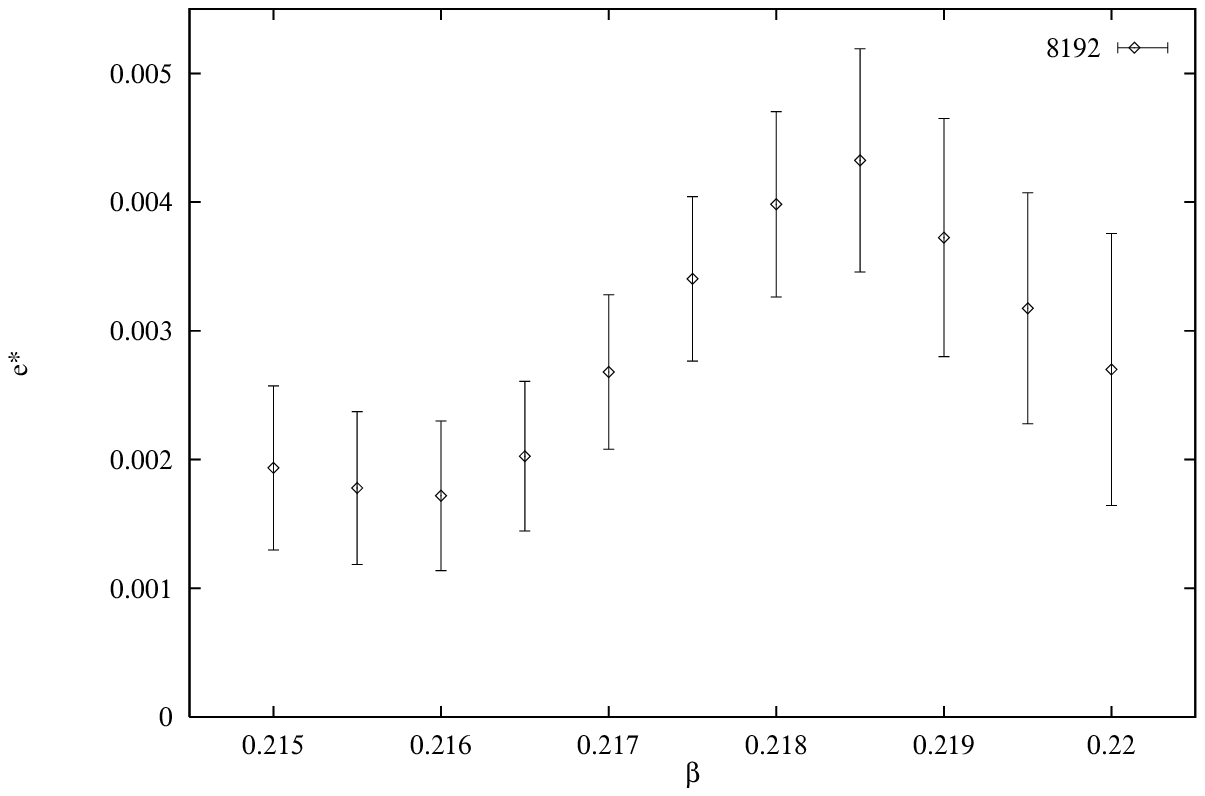}
\protect\caption{\protect\label{enecorrfig}
The correlation $e^*$ between the energies of different species
 plotted for the $c=1$ model on
$N=8192$ lattices.}
\end{figure}

        We finally turn to an examination of correlations (defined in
(\ref{magcorr}) and (\ref{enecorr})) between the different spin
species.  Fig.~\ref{magcorrfig} exhibits definite, though moderately
small, correlations between the spins of different species.
As evident in Fig.~\ref{enecorrfig}, the correlations between the
average species' energies, as measured by $e^*$, are much smaller.
The magnetization (and more directly, the susceptibility) is strongly
correlated with the distribution of FK cluster sizes, which in turn
should be sensitive to the bottlenecks which characterize the
worldsheet geometry.  Since the correlations between species are
mediated by fluctuations in the geometry, it is not surprising that
they are stronger in the magnetic sector than in the energy sector.
In both cases the correlations are not so strong, indicating that the
Ising spins are only weakly coupling to gravity for the lattice sizes
we consider.

\section{Discussion \protect\label{Disc}}

    To gauge our prospects for extending this work to $c>1$, we shall
now attempt to shed more light on the origin of the corrections to
scaling at $c=1$.  The dynamics at $c=1$ is governed by the tachyon
(the lowest mass operator) which acquires a continuous spectrum with
vanishing energy at $0$ momentum.  The scaling corrections arise from
the infrared cutoff on the tachyonic momentum, $p > 1/\ln N$
\cite{KLE}.  The origin of this logarithmic cutoff may not be so
evident in the context of the two-species Ising model, since the
tachyon cannot be expressed so simply in terms of the Ising spins.
The presence of logarithmic corrections appears to only depend on the
value of the central charge, so we appeal to universality and consider
instead the $c=1$ Gaussian theory.  The tachyon can then be written in
terms of the Gaussian field $X$ as
\begin{equation}
 T(p) \sim \int \sqrt{|{\hat{g}}|}\exp (ipX + (|p|-2)\phi);
\end{equation}
$\phi$ is the Liouville field and ${\hat{g}}$ is the reference metric
($g = {\hat{g}}\exp(-\phi)$).   The infrared target space momentum
cutoff is then just a consequence of the infinite Hausdorff dimension
of the embedding space \cite{KazMig},
\begin{equation}
\langle X X \rangle \sim (\ln N)^2.
\end{equation}
The suppression of tachyon propagation by finite-size effects
effectively weakens the coupling between gravity and the Ising spins.
Therefore, the spins of each species are only very weakly coupled, and
the two and one species models appear to be qualitatively very similar
on the lattices we consider.  This should not be true in the continuum
limit. We see that without an understanding of the corrections to
scaling for $c=1$, the numerical observations seem to conflict with
our theoretical expectations of the onset of strong coupling at $c=1$.

       The target space Hausdorff dimension, as measured for Gaussian
embeddings of $d$ somewhat greater than $1$, is also large.  Thus we
would expect that for $c>1$, the tachyon ground state energy still
acquires a considerable shift and we expect large corrections to
scaling. At $c=1$, we were able to determine these corrections because
the $c=1$ model is solvable.  Without new theoretical input, the
prognosis for understanding simulations for $c$ somewhat greater than
$1$ thus seems poor.

\vfill\eject

\section{Acknowledgments}

This work has been done with NPAC (Northeast Parallel Architectures
Center) computing facilities.  We would like to thank John
Apostolakis, Simon Catterall and Paul Coddington for helpful
correspondence and conversations.  The research of MB was supported by
the Department of Energy Outstanding Junior Investigator Grant DOE
DE-FG02-85ER40231, that of MF by funds from NPAC and that of GH by
research funds from Syracuse University.

\vfill






\end{document}